\documentclass[aps,pra,twocolumn,superscriptaddress,groupedaddress]{revtex4}  
\usepackage{graphicx}  
\usepackage{dcolumn}   
\usepackage{bm}        
\usepackage{amssymb}   
\usepackage{upgreek}
\usepackage{amsmath}
\usepackage{braket}
\usepackage{color}

\begin{document}


\title{Electro-optic polarization tuning of microcavities with a single quantum dot}

\author{J. A. Frey}
\email{jfrey@physics.ucsb.edu}
\affiliation{Department of Physics, University of California, Santa Barbara, California 93106, USA}
\author{H. J. Snijders}
\affiliation{Huygens-Kamerlingh Onnes Laboratorium, Universiteit Leiden, 2333 CA Leiden, The Netherlands}
\author{J. Norman}
\affiliation{Department of Electrical $\&$ Computer Engineering, University of California, Santa Barbara, California 93106, USA}
\author{A. C. Gossard}
\affiliation{Department of Electrical $\&$ Computer Engineering, University of California, Santa Barbara, California 93106, USA}
\author{J. E. Bowers}
\affiliation{Department of Electrical $\&$ Computer Engineering, University of California, Santa Barbara, California 93106, USA}
\author{W. L\"{o}ffler}
\affiliation{Huygens-Kamerlingh Onnes Laboratorium, Universiteit Leiden, 2333 CA Leiden, The Netherlands}
\author{D. Bouwmeester}
\affiliation{Department of Physics, University of California, Santa Barbara, California 93106, USA}
\affiliation{Huygens-Kamerlingh Onnes Laboratorium, Universiteit Leiden, 2333 CA Leiden, The Netherlands}

\date{\today}

\begin{abstract}
We present an oxide aperture microcavity with embedded quantum dots that utilizes a three contact design to independently tune the quantum dot wavelength and birefringence of the cavity modes. A polarization splitting tuning of $\sim$5 GHz is observed. For typical microcavity polarization splittings, the method can be used to achieve perfect polarization degeneracy that is required for many polarization-based implementations of photonic quantum gates. The embedded quantum dot wavelength can be tuned into resonance with the cavity, independent of the polarization tuning.
\end{abstract}

\maketitle 


Semiconductor quantum dots (QDs) are promising candidates for various aspects of quantum information, such as single photon sources \cite{Somaschi2016}, remote-spin entanglement generation \cite{Delteil2016, Stockill2017, Wei2014}, generation of photonic cluster states \cite{Lindner2009}, and direct production of entangled photon pairs \cite{Benson2000, Stevenson2006, Muller2014, Trotta2014}. Scalable architectures, however, require a high collection efficiency of the emitted photons. For this reason it is necessary to embed the QDs in an optical microcavity to enhance emission into a single desired mode \cite {Senellart2017}. With the advances of cavity designs, high quality microcavities with embedded QDs can be routinely fabricated. Purcell enhancement \cite{Gerard1998} and strong coupling \cite{Reithmaier2004} of a single QD in a micropillar cavity have been demonstrated, and quality factors exceeding 250,000 have been reported \cite{Schneider2016}. The inevitable birefringence of such structures, however, presents an additional complication with ever increasing quality factors. For the implementation of polarization based quantum gates, it is necessary for an embedded QD to couple equally to both photon polarizations, and therefore to both orthogonally polarized cavity modes. Single photon sources often require resonant pulsed excitation, where it is impossible to eliminate all incident laser light with cross polarization if the cavity is birefringent over the spectral width of the laser.

Inherent material birefringence does not exist for the zinc-blende crystal structure. Polarization splitting of cavity modes arises primarily from electro-optic (EO) or elasto-optic birefringnece, and form birefringence from asymmetries of the fabricated cavity. QDs are typically embedded in a diode in order to tune their emission wavelength through the quantum confined Stark effect. Because of the linear EO effect, or Pockels effect, the applied field introduces birefringence. The built-in electric fields and strain in a heterostructure, either inherent from growth or the result of fabrication, also contribute to the polarization splitting of the cavity modes \cite{VanExter1997, Hendriks1997}. EO birefringence is the likely reason micropillar cavities often exhibit linearly polarized modes aligned with the major and minor axes of the crystal. Asymmetrical cavity shapes further contribute to polarization splitting, see \cite{Bava2001, Debernardi2002} for example. There has been a lot of work in fabricating microcavities with close to zero polarization splitting \cite{Bakker2015, Bonato2009, Bakker2014}, but a continuous fine tuning method is still needed.

\begin{figure}
	\includegraphics[width=\linewidth]{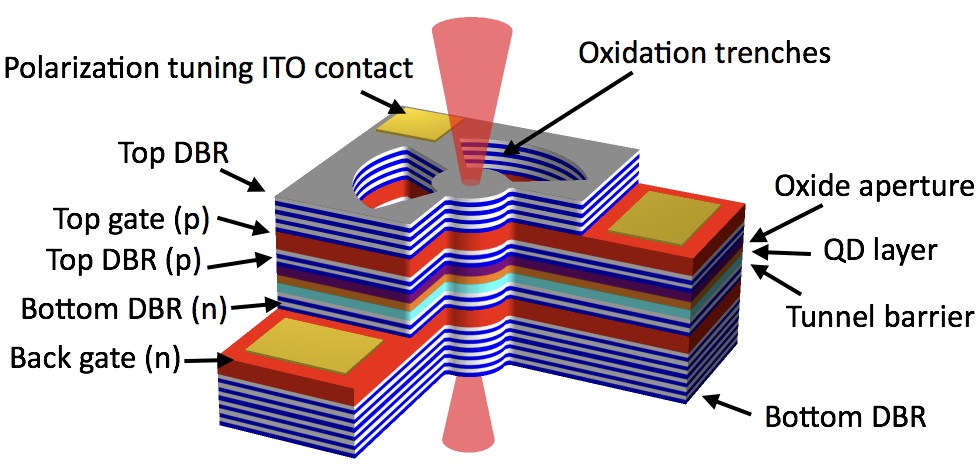}
	\caption{Representation of the microcavity. A p-i-n doped active region around the QDs controls their emission wavelength, and an ITO top contact can be biased with respect to the p layer to separately apply a voltage to the top mirror.}
	\label{fig:microcavity}
\end{figure}

The device presented, shown in Figure \ref{fig:microcavity}, is a micropillar cavity with three contacts. The QDs are embedded in a p-i-n diode and surrounded by two distributed Bragg reflectors (DBRs). A transparent indium tin oxide (ITO) Schottky contact is deposited on the top of the device. Applying an electric field over the top mirror tunes the cavity polarization splitting through the Pockels effect. The electric field in the active region can be set independently to tune the QD into resonance with the cavity mode. Such a device has similarly been used to control the output polarization state of a vertical-cavity surface-emitting laser through modification of the optical gain of the polarized modes \cite{Park2000}. With polarization resolved reflection spectroscopy, we study the EO tuning on the scale of GHz and demonstrate complete restoration of the cavity polarization splitting. Electrical control of the embedded QD wavelength is retained while the cavity polarization splitting is tuned.

The Pockels effect is a second order effect in crystals that lack an inversion center \cite{Abeles1972}. For crystals in the $\bar{4}$3m point group, including zinc-blende crystals, the linear electro-optic tensor has only 1 non-zero element, $r_{41}$, assumed here to be real \cite{Adachi1984}. For an electric field applied along the $\left[001\right]$ axis, the crystal is bireferingent, and the corresponding principle axes of the crystal are defined by the major $\left[110\right]$ and minor $\left[1\bar{1}0\right]$ axes. The induced birefringence from an applied DC field, $E_{dc}$, is described by
\begin{equation}\label{eq:pockels}
	\frac{1}{n_{\pm}^{2}}-\frac{1}{n_{0}^{2}}=\pm r_{41}E_{dc}
\end{equation} 
where $n_{0}$ is the real index with no field and $n_{\pm}$ refers to the real index of the $\left[110\right] \left(+\right)$ and $\left[1\bar{1}0\right]\left(-\right)$ axes. 

 
DBR mirrors have a large penetration depth, which allows a cavity resonance to be tuned by applying an electric field only over a single mirror. It is well known that an incident optical pulse undergoes a time delay and spreading in space upon reflection off of a dispersive mirror \cite{Babic1992,Laporta1985}. The time delay is important for characterizing the wavelength of a cavity, which to first order is defined by a linear phase slope, $\frac{d\phi}{d\omega}\Big|_{\omega=\omega_{0}}$. The effective penetration depth of a \textit{hard} mirror, $L_{eff}$, is given by
\begin{equation}\label{eq:pendepth}
	L_{eff}=-\frac{c}{2n_{1}}\frac{d\phi}{d\omega}\Bigg|_{\omega=\omega_{0}}.
\end{equation}
Equation \ref{eq:pendepth} is evaluated at the Bragg condition, denoted $\omega = \omega_{0}$. $L_{eff}$ is used to model a dispersive mirror with a fixed phase mirror placed some distance from the actual mirror interface. In general, it is irrelevant what material is chosen to replace the resulting empty space, so long as the optical length is kept the same. For calculations involving cavities, it is most convenient to choose the material in front of the mirror to be the same material composing the spacer region of the cavity, denoted by $n_{1}$. Analytical solutions for the penetration depth exist in the case of a single DBR mirror. For details the reader is referred to the literature \cite{Babic1992, Laporta1985, Coldren, Corzine1991}.

The optical path length of the layers in a DBR are altered from the Pockels effect when a field is present; the Bragg frequency shifts as a result of an applied voltage. In the effective mirror model, the shift is equivalent to moving the mirror some distance, $\Delta L_{eff}$,
\begin{equation}\label{eq:pendepthchange}
	\Delta L_{eff}=\frac{2n_{1}}{\omega_{0}n_{1,gr}}L_{eff}\Delta\omega_{Bragg}.
\end{equation}
$\Delta\omega_{Bragg}$ is the frequency shift of the Bragg condition and $n_{1,gr}$ is the group index of the spacer material. Considering a Fabry-Perot cavity formed from two identical DBR mirrors, described in the effective mirror model, with penetration depths given by Equation \ref{eq:pendepth}, and a spacer layer of length, $ L_{spacer}$, the cavity resonance shift can be calculated. The frequency shift of the cavity, $\Delta\omega$, is given by
\begin{equation}\label{eq:fractwavechange}
	\Delta\omega=\frac{n_{1}}{n_{1,gr}}\frac{2L_{eff}}{L_{spacer}+2L_{eff}}\Delta\omega_{Bragg}.
\end{equation}
Equation \ref{eq:fractwavechange} assumes that the second mirror is unaltered by the application of an electric field on the other, which is only true if the second mirror is non-dispersive. For two DBR mirrors, the cavity shift from the change in penetration depth of one mirror is accompanied by a change in penetration depth of the second. The coupling of both mirrors with a shift in cavity frequency must be taken into account. The parameter $k=\frac{n_{1}}{n_{1,gr}}\frac{2L_{eff}}{L_{spacer}+2L_{eff}}$ is introduced for simplicity. The total change in cavity resonance with coupled mirrors is given by adding the contributions of both mirrors,
\begin{equation}\label{eq:symmetriccavityshift}
	\Delta\omega = \frac{k}{1 + 2k}\Delta\omega_{Bragg}.
\end{equation}
$\Delta\omega_{Bragg}$ is found from computing the new optical path length of a single period in the DBR with an applied field. The cavity polarization splitting is denoted $\Delta\omega^{\prime}$. For small changes in the index of refraction of both materials composing the DBR cavity, $n_{1}$ and $n_{2}$, the polarization splitting can be expressed as
\begin{equation}\label{eq:totalbirefringence}
	\Delta\omega^{\prime}=\frac{k}{1 + 2k}\frac{\omega_{0}}{2}\left(\frac{\Delta n_{1}}{n_{1}} +\frac{\Delta n_{2}}{n_{2}}\right).
\end{equation}
The values $\Delta n_{1}$ and $\Delta n_{2}$ are the changes in refractive index for materials 1 and 2.

Equation \ref{eq:totalbirefringence} assumes two identical mirrors with a cavity resonance matching exactly the Bragg condition. Our devices have more complex designs than this simplified model. Nevertheless, the model can give a rough estimate of the expected tuning range with an applied voltage over a single mirror, and yield insight into the effects of various parameters that may be exploited to increase the tuning range.

Samples are grown with molecular beam epitaxy on an undoped $\left[001\right]$ GaAs substrate. Two DBR mirrors, composed of quarter wave layers (24 bottom layers and 22 top layers) of GaAs and AlGaAs, surrounding the active region form a Fabry-Perot cavity. InGaAs QDs are grown via Stranski-Krastanov growth in the active region. A large part of the DBR layers is undoped, and a 60 nm layer of ITO is deposited on top of the device to form a transparent Schottky contact. Microcavities with transverse confinement are fabricated by etching trenches with Cl$_{2}$/BCl$_{3}$/N$_{2}$ inductively coupled plasma. Etching both exposes a high Al concentration layer, and provides access to the n and p doped contact layers. The sample is held in a tube furnace at $420^{\circ}$C in steam for $\sim$1 hour for the oxidation process. The Al concentration is tapered along the oxide layer in order to create an intracavity lens \cite{Coldren1998}. Ni/AuGe/Ag/Au and Ti/Pt/Au contact pads are deposited on the n and p doped contact layers respectively to form Ohmic contacts. The ITO layer is contacted via a Cr/Ni/Au pad, and all three pads are wire bonded to a custom submount before mounting in a closed cycle cryostat operating at 6K. 

Asymmetric apertures are likely the primary cause of initial birefringence in oxide aperture microcavities. In order to completely restore polarization degeneracy, the shape of the aperture must be controlled. The mesa in this study is made elliptical such that it is longer along the major axis to account for the difference in oxidation rates along the major and minor axes \cite{Choquette1996, Bakker2014}. By altering the aperture symmetry, the polarization can be coarsely tuned to degeneracy and EO tuning is used as a fine control.

\begin{figure}
	\includegraphics[width=\linewidth]{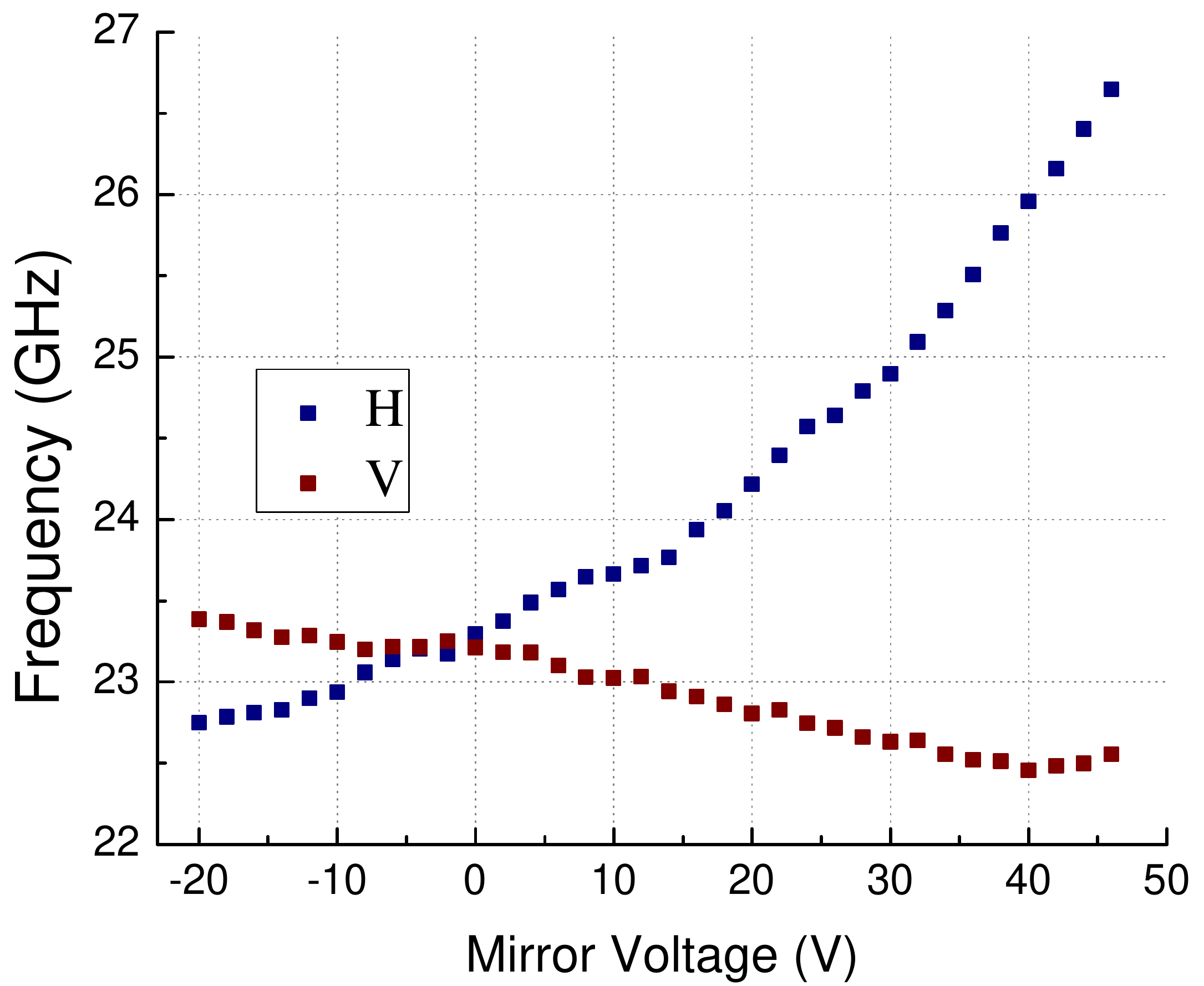}
	\caption{Typical tuning behavior of the cavity modes as a function of the applied field. There is a clear crossing of the two modes indicating the presence of a perfectly degenerate configuration. There is an absolute shift of both modes which can be explained by EO effects that do not induce birefringence.}
	\label{fig:fielddependance}
\end{figure}

Reflection spectroscopy on the lowest order HE$_{11}$ Gaussian modes of the cavity reveals in detail the behavior of the modes with EO tuning. Figure \ref{fig:fielddependance} shows the typical tuning behavior of the modes, denoted H and V, as a bias is applied to the top contact. The p-i-n region is held at a fixed voltage with respect to the p contact. H corresponds to the major axis of the crystal, $\left[110\right]$. The modes cross around -5 V implying the existence of a configuration that gives a completely polarization degenerate cavity. In addition, there is an absolute shift of the cavity resonance, which can be explained by quadratic EO effects that do not exhibit birefringence: Kerr effect, etc.

The tuning range observed experimentally is aproximately 4.5 GHz, which is non-destructive and reversible. Parameters derived from the device design yield a penetration depth, $L_{eff}=382$ nm, and the spacer layer thickness is, $L_{spacer}=536$ nm. Values for the linear EO coefficient are taken to be $r_{41}=1.72\:x\:10^{-10}\frac{\mathrm{cm}}{\mathrm{V}}$ and $r_{41}=1.00\:x\:10^{-10}\frac{\mathrm{cm}}{\mathrm{V}}$ for GaAs and AlGaAs respectively \cite{Berseth1992}. An estimate of the tuning range using equation \ref{eq:totalbirefringence} then yields $\Delta\omega^{\prime}=2.2$ GHz. The theoretical model is verified with transmission matrix method (TMM) calulations. TMM on a simplified symmetric cavity yields a tuning range of 2.2 GHz. The TMM calculation for the full device structure yields a tuning range of 0.73 GHz, about 5 times less than observed in the experiment.

Several effects could explain the deviation of the observed shift from theory. In addition to the deviation from theory we observe a $\sim10\:\%$ increase in the width of the H mode at high fields but not for the V mode. The device layout has contacts oriented along the major axis and small transverse currents could introduce free carrier losses for only the H mode. Although mirror absorption can alter the phase dispersion, the observed increase in width does not correspond to values of absorption that are expected to result in measurable shifts of the cavity resonance. For transverse fields applied to the device, quadratic EO effects can induce birefringence. Typical values of the quadratic EO coefficient for GaAs are small, but very dispersive around the band gap \cite{Berseth1992}. Large quadratic effects have additionally been observed in complicated waveguide heterostructures \cite{Glick1986, Wood1987}.

\begin{figure}
	\includegraphics[width=\linewidth]{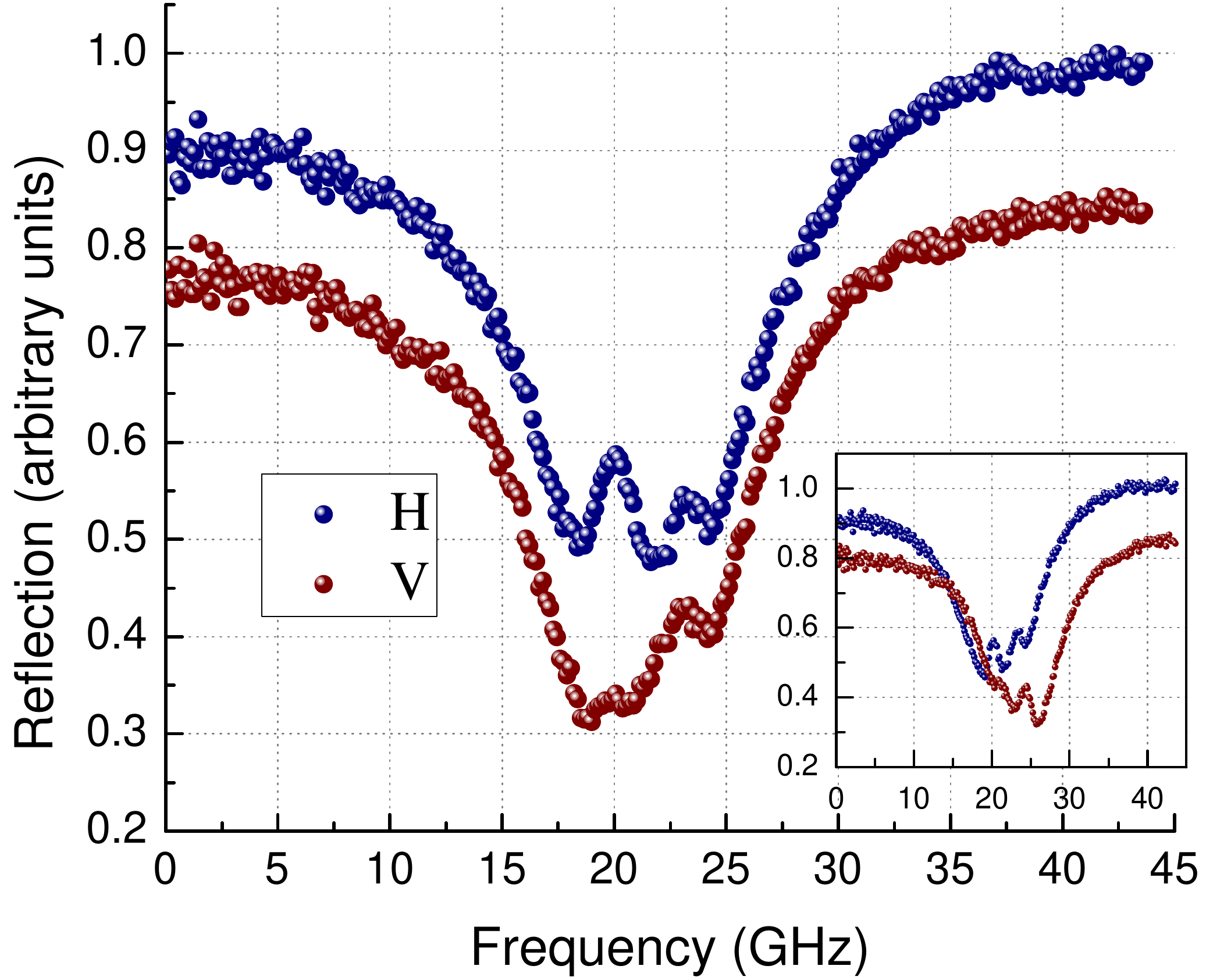}
	\caption{Polarization resolved reflection spectra for a polarization degenerate cavity. The incident laser light is aligned along the cavity axis H and V. The two exciton transitions are visible as peaks in the reflection dip. The inset shows the same measurement with the cavity tuned away from degeneracy. H and V are offset for clarity.} 
	\label{fig:insetgraph}
\end{figure}

The QD-microcavity system must be excited in the cavity polarization basis when the polarization degeneracy is large. An example of this, albeit with small polarization splitting, is shown in Figure \ref{fig:insetgraph}. H and V denote the cavity basis. In the reflection spectrum of the cavity, the QD appears as a peak in the cavity mode, which is due to the excitation of an electron-hole pair of a neutral dot. The fine structure splitting of the QD is clearly visible as two peaks. The inset shows the same measurement for the case of imperfect polarization degeneracy.

\begin{figure}
	\includegraphics[width=\linewidth]{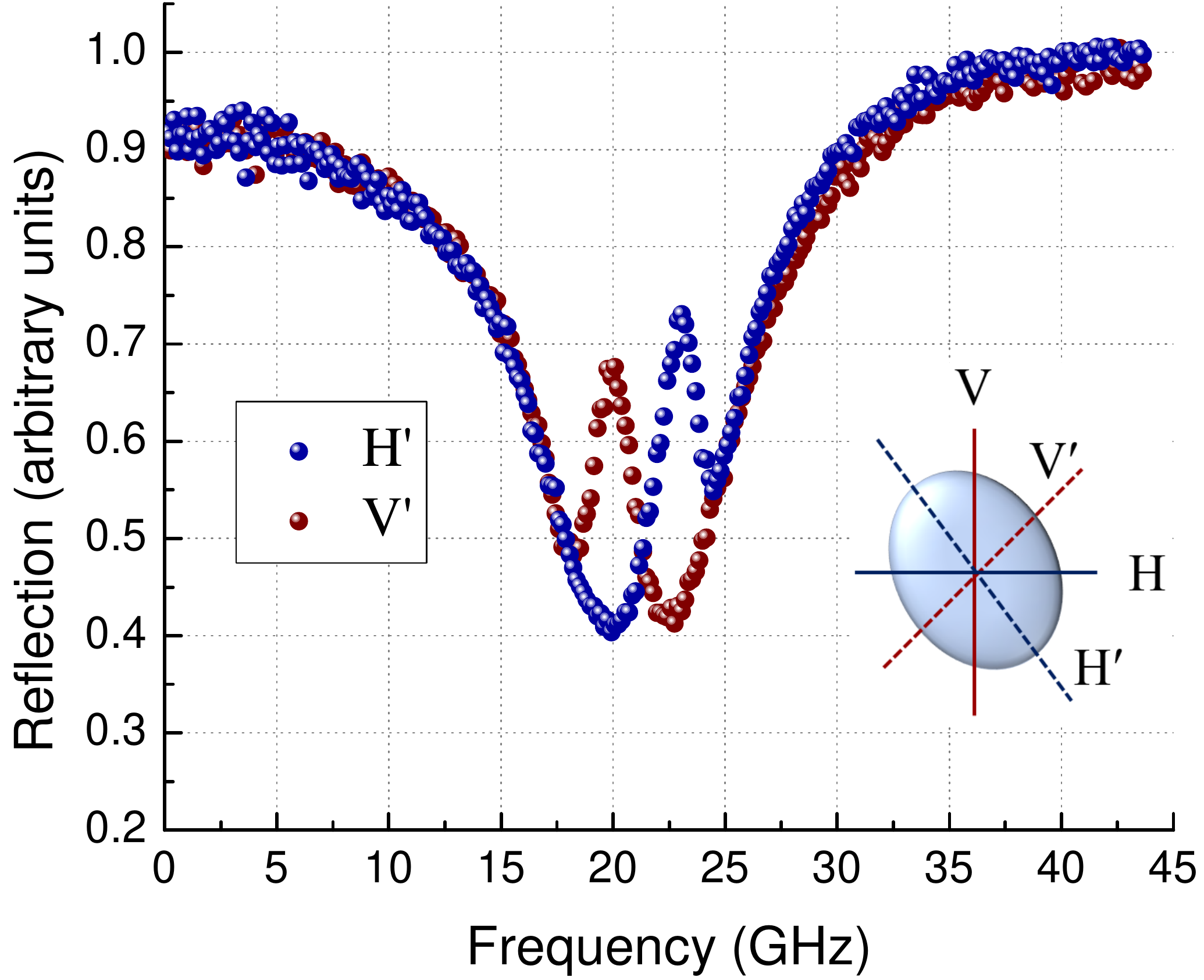}
	\caption{Reflection spectra of a completely polarization degenerate cavity in the basis of the neutral exciton transitions. Maximum coupling to either transition can be achieved only if the cavity has no birefringence.}
	\label{fig:poldegdotbasis}
\end{figure}

Self-assembled InGaAs QDs grown on  $\left[001\right]$ substrates are typically not oriented along either the major or minor crystal axis. The resulting cavity-QD system, therefore, has a cavity polarization basis and QD polarization basis. Maximum coupling of photons to exciton transitions can be accomplished in the case of identical cavity and QD basis, or when the cavity is polarization degenerate. The latter is shown in Figure \ref{fig:poldegdotbasis}, which shows the reflection spectrum in the QD basis, denoted as H$^{\prime}$ and V$^{\prime}$, approximately 45$^{\circ}$ from the cavity basis. For either exciton transition, the coupling to the cavity mode is stronger than in the cavity basis, evident by the higher peaks in reflectivity. 

State of the art single photon sources require pulsed excitation \cite{Senellart2017} for which the light uncoupled to a QD must be extinguished. Cavity polarization splitting, however, results in spectrally dependent rotations of uncoupled incident laser light, which limits the purity of a single photon source. The presence of a cavity basis and exciton basis further restricts the possible polarization angles for efficient single photon emission into a cavity mode \cite{Anton2017}. These issues can be remedied only with a polarization degenerate cavity.

We have demonstrated that a three contact microcavity design gives an extra knob to control the polarization splitting of microcavities, and allows simultaneous electrical control of the embedded QDs. The method can be extended to more conventional air guided micropillar cavities \cite{Somaschi2016}, and/or combined with aditional tuning techniques, such as strain, which can address the fine structure splitting of the dots. The fabrication process is compatible with methods of identifying particular dots prior to fabrication that deterministically control the QD properties and coupling to the cavity \cite{Dousse2008}. The QD studied here is a neutral dot. For charged QDs, which exhibit circularly polarized transitions, polarization degeneracy of the cavity is even more important. Several proposals are available for trapping single electrons or spins \cite{Heiss2009,Mar2014}, and electron spins have been demonstrated in photonic crystal cavities \cite{Carter2013}. Together with polarization degenerate cavities, this system would have all the necessary ingredients for efficient photonic quantum gates and spin-photon entanglement schemes.

\bigskip
\noindent
\textbf{Funding.} Foundation for Fundamental Research on Matter-Netherlands Organisation for Scientific Research (FOM-NWO) (08QIP6-2) as part of the Frontiers of Nanoscience program; National Science Foundation (NSF) (0901886, 0960331).

\bigskip
\noindent
\textbf{Acknowledgments.} The authors would like to acknowledge A. Kerr for sample characterization, and B. J. Thibeault and D. D. John for fruitful discussions.
\bigskip
\noindent 

\bibliographystyle{h-physrev}
\bibliography{EOTuningArXiv}

\begin{thebibliography}{10}

\bibitem{Somaschi2016}
N.~Somaschi {\em et~al.},
\newblock Nature Photonics {\bf 10}, 340 (2016).

\bibitem{Delteil2016}
A.~Delteil {\em et~al.},
\newblock Nature Physics {\bf 12}, 218 (2016).

\bibitem{Stockill2017}
R.~Stockill {\em et~al.},
\newblock Physical Review Letters {\bf 119}, 010503 (2017).

\bibitem{Wei2014}
H.-R. Wei and F.-G. Deng,
\newblock Optics Express {\bf 22}, 593 (2014).

\bibitem{Lindner2009}
N.~H. Lindner and T.~Rudolph,
\newblock Physical Review Letters {\bf 103}, 113602 (2009).

\bibitem{Benson2000}
O.~Benson, C.~Santori, M.~Pelton, and Y.~Yamamoto,
\newblock Physical Review Letters {\bf 84}, 2513 (2000).

\bibitem{Stevenson2006}
R.~M. Stevenson {\em et~al.},
\newblock Nature {\bf 439}, 179 (2006).

\bibitem{Muller2014}
M.~M{\"{u}}ller, S.~Bounouar, K.~D. J{\"{o}}ns, M.~Gl{\"{a}}ssl, and
  P.~Michler,
\newblock Nature Photonics {\bf 8}, 224 (2014).

\bibitem{Trotta2014}
R.~Trotta, J.~S. Wildmann, E.~Zallo, O.~G. Schmidt, and A.~Rastelli,
\newblock Nano Letters {\bf 14}, 3439 (2014).

\bibitem{Senellart2017}
P.~Senellart, G.~Solomon, and A.~White,
\newblock Nature Nanotechnology {\bf 12}, 1026 (2017).

\bibitem{Gerard1998}
J.~G{\'{e}}rard {\em et~al.},
\newblock Physical Review Letters {\bf 81}, 1110 (1998).

\bibitem{Reithmaier2004}
J.~P. Reithmaier {\em et~al.},
\newblock Nature {\bf 432}, 197 (2004).

\bibitem{Schneider2016}
C.~Schneider, P.~Gold, S.~Reitzenstein, S.~H{\"{o}}fling, and M.~Kamp,
\newblock Applied Physics B {\bf 122}, 19 (2016).

\bibitem{VanExter1997}
M.~P. van Exter, A.~K. {Jansen van Doorn}, and J.~P. Woerdman,
\newblock Physical Review A {\bf 56}, 845 (1997).

\bibitem{Hendriks1997}
R.~F.~M. Hendriks {\em et~al.},
\newblock Applied Physics Letters {\bf 71}, 2599 (1997).

\bibitem{Bava2001}
G.~P. Bava, P.~Debernardi, and L.~Fratta,
\newblock Physical Review A {\bf 63}, 023816 (2001).

\bibitem{Debernardi2002}
P.~Debernardi, G.~Bava, C.~Degen, I.~Fischer, and W.~Elsasser,
\newblock IEEE Journal of Quantum Electronics {\bf 38}, 73 (2002).

\bibitem{Bakker2015}
M.~P. Bakker {\em et~al.},
\newblock Physical Review B {\bf 91}, 115319 (2015).

\bibitem{Bonato2009}
C.~Bonato {\em et~al.},
\newblock Applied Physics Letters {\bf 95}, 251104 (2009).

\bibitem{Bakker2014}
M.~P. Bakker {\em et~al.},
\newblock Applied Physics Letters {\bf 104}, 151109 (2014).

\bibitem{Park2000}
M.~S. Park {\em et~al.},
\newblock Applied Physics Letters {\bf 76}, 813 (2000).

\bibitem{Abeles1972}
F.~Abeles,
\newblock {\em {Optical Properties of Solids}} (, 1972).

\bibitem{Adachi1984}
S.~Adachi and K.~Oe,
\newblock Journal of Applied Physics {\bf 56} (1984).

\bibitem{Babic1992}
D.~Babic and S.~Corzine,
\newblock IEEE Journal of Quantum Electronics {\bf 28}, 514 (1992).

\bibitem{Laporta1985}
P.~Laporta and V.~Magni,
\newblock Applied Optics {\bf 24}, 2014 (1985).

\bibitem{Coldren}
L.~A. Coldren, S.~W. Corzine, and M.~L. Milan,
\newblock {\em {Diode Lasers and Photonic Integrated Circuits}}, Second ed. .

\bibitem{Corzine1991}
S.~Corzine, R.~Yan, and L.~Coldren,
\newblock IEEE Journal of Quantum Electronics {\bf 27}, 2086 (1991).

\bibitem{Coldren1998}
L.~A. Coldren, B.~J. Thibeault, E.~R. Hegblom, G.~B. Thompson, and J.~W. Scott,
\newblock Applied Physics Letters {\bf 68}, 313 (1996).

\bibitem{Choquette1996}
K.~D. Choquette {\em et~al.},
\newblock Applied Physics Letters {\bf 69}, 1385 (1996).

\bibitem{Berseth1992}
C.-A. Berseth, C.~Wuethrich, and F.~K. Reinhart,
\newblock Journal of Applied Physics {\bf 71} (1992).

\bibitem{Glick1986}
M.~Glick, F.~K. Reinhart, G.~Weimann, and W.~Schlapp,
\newblock Applied Physics Letters {\bf 48}, 989 (1986).

\bibitem{Wood1987}
T.~H. Wood, R.~W. Tkach, and A.~R. Chraplyvy,
\newblock Applied Physics Letters {\bf 50}, 798 (1987).

\bibitem{Anton2017}
C.~Ant{\'{o}}n {\em et~al.},
\newblock Optica {\bf 4}, 1326 (2017).

\bibitem{Dousse2008}
A.~Dousse {\em et~al.},
\newblock Physical Review Letters {\bf 101}, 267404 (2008).

\bibitem{Heiss2009}
D.~Heiss {\em et~al.},
\newblock Applied Physics Letters {\bf 94}, 072108 (2009).

\bibitem{Mar2014}
J.~D. Mar, J.~J. Baumberg, X.~Xu, A.~C. Irvine, and D.~A. Williams,
\newblock Physical Review B {\bf 90} (2014).

\bibitem{Carter2013}
S.~G. Carter {\em et~al.},
\newblock Nature Photonics {\bf 7}, 329 (2013).

\end{thebibliography}

\end{document}